\documentclass[a4,twocolumn,superscriptaddress,showpacs,preprintnumbers,amsmath,amssymb]{revtex4}

\usepackage{amsmath}
\usepackage{graphicx}
\usepackage{dcolumn}

\begin{document}

\title{Two-component mixture of charged particles confined in a channel: melting}
\author{W.~P.~Ferreira}
\email{wandemberg@fisica.ufc.br} \affiliation{Departamento de
F\'isica, Universidade Federal do Cear\'a, Caixa Postal 6030,
Campus do Pici, 60455-760
Fortaleza, Cear\'a, Brazil}%
\author{G.~A.~Farias}
\email{gil@fisica.ufc.br} \affiliation{Departamento de F\'isica,
 Universidade Federal do Cear\'a, Caixa Postal 6030, Campus do
 Pici, 60455-760 Fortaleza, Cear\'a, Brazil}%
\author{F.~M.~Peeters}
\email{francois.peeters@ua.ac.be} \affiliation{Departamento de
F\'isica, Universidade Federal do Cear\'a, Caixa Postal 6030,
Campus do Pici, 60455-760 Fortaleza, Cear\'a, Brazil}
\affiliation{Department of Physics, University of Antwerp,
Groenenborgerlaan 171, B-2020 Antwerpen, Belgium}%

\date{ \today }

\begin{abstract}
The melting of a binary system of charged particles confined in a
{\it quasi}-one-dimensional parabolic channel is studied through
Monte Carlo simulations. At zero temperature the particles are
ordered in parallel chains. The melting is anisotropic and
different melting temperatures are obtained according to the
spatial direction, and the different types of particles present in
the system. Melting is very different for the single-, two- and
four-chain configurations. A temperature induced structural phase
transition is found between two different four chain ordered
states which is absent in the mono-disperse system. In the mixed
regime, where the two types of particles are only slightly
different, melting is almost isotropic and a thermally induced
homogeneous distribution of the distinct types of charges is
observed.
\end{abstract}


\maketitle

\section{Introduction}
\label{sec:introduction}

Recently, there has been a renewed interest in the study of the
physical properties of multi-component systems
\cite{liu06a,shevchenko06,leunissen05}. The presence of particles
with distinct physical properties (e.g. size, charge, mass)
introduces a competition between different scales, which is the
reason for the richer phenomenology in such systems \cite{liu06a}.
The simplest multi-component system is a binary mixture of two
types of particles which, as compared to a mono-disperse system
\cite{kaufman_apl2009,overduin_epl2009,assoud07,wand05,wand06,gio04,wandprb08},
has been shown to exhibit a very rich phase diagram. Such binary
mixtures have been studied in several different experimental
setups such as ion traps \cite{levi88, drewsen98}, magnetized
disks floating at a liquid-air interface in an external magnetic
field \cite{grzybowski01}, dusty plasmas \cite{psakhie08} and
colloidal suspensions \cite{hoffmann06}.

Under specific conditions of density and temperature a system of
interacting particles will solidify into an ordered arrangement
forming a crystal structure. When the inter-particle potential is
purely repulsive this is often referred to as a Wigner crystal, in
reference to the crystallization of the electron gas predicted by
Eugene Wigner \cite{wigner34}, and observed experimentally for the
first time in a 2D system of electrons on the surface of liquid
helium by Grimes and Adams \cite{grimes1979}. In three-dimensional
case (3D), a first experimental investigation of classical plasma
crystals can be found in Ref. \cite{arp2004}. Such an ordered
phase appears in several non-electronic systems, and has also been
experimentally observed, e.g. in ion traps \cite{levi88,
drewsen98}, colloidal systems \cite{zahn99,golosovsky02}, and
dusty plasmas \cite{chu1994,liu03,liu05}. Besides the
single-component system, which has been widely studied in the last
years, crystallization may also be observed in multi-component
systems of interacting particles \cite{liu06a,leunissen05}, and
particularly in binary systems \cite{wand05}. Recently, a large
number of different equilibrium configurations, which depend on
the relative fraction of the different types of particles, was
found in a two-dimensional (2D) system of interacting dipoles
\cite{assoud07}.

In the case of a $quasi$-one-dimensional (Q1D) system of repulsive
particles several recent experimental
\cite{liu03,liu05,haghgooie06,koppl06}, analytical and numerical
\cite{haghgooie04,gio04,wandprb08} studies showed interesting
physical properties and promising perspectives for practical
applications \cite{doyle02}. Systematic studies of Q1D systems of
interacting particles confined in parabolic\cite{gio04} and
hardwall \cite{haghgooie04,yang09} channels revealed many
different phases of chain-like structures, and even poorly ordered
configurations. The latter depends on the nature of the
confinement potential and is favored by hard wall-type of
confinement.

An interesting and very important physical phenomenon is melting.
In the 2D case, the transition from a solid phase to a complete
isotropic liquid phase can be characterized by a two-step process
with an intermediate hexatic phase
\cite{thouless73,halperin78,young79}. However, such a melting
scenario is not unique and depends on the considered system, i.e.
the inter-particle interaction, and its dimensionality.
Furthermore, previous papers showed also that the melting of 2D
clusters is non-universal \cite{zahn99,branicio,bedanov94}.

The specific structural arrangement of the particles in confined
systems has a very important effect on the nature and specific
process of melting. In the case of 2D  mono- and bi-disperse
confined clusters, structural properties of the ordered configuration were found to strongly
influence the melting \cite{bedanov94,wand06}. E.g., the melting
temperature in small clusters of charged particles was found to be a nonuniform increasing
function of the strength of the binding of the cluster
\cite{wandpre05}. It was demonstrated in Ref.
[\onlinecite{wandpre05}] that the symmetry of the configuration is
one of the dominant factors that determines the melting
temperature. In the case of small confined binary clusters, we
showed recently that thermally induced structural phase
transitions can occur before melting takes place. In addition, a
remarkable temperature induced spatial phase separation was found
\cite{wand06}.

In the present paper we use Monte Carlo simulations to study the
melting of a binary system of classical charged particles confined
in a  {\it quasi}-one-dimensional parabolic channel. The structure
and normal mode spectrum of such a system were recently addressed
by us in Ref. [\onlinecite{wandprb08}]. A very rich set of
chain-like structures with, in some cases, an intrinsic spatial
separation of the distinct types of particles was predicted. A
density dependent band-gap in the phonon spectrum was found, and a
softening of the normal mode frequencies characteristic of a
continuous structural transition from the mixed single-chain to
the segregated two-chain regime. In the present analysis we limit
ourselves to an equal density of the two types of particles which
are assumed to have the same mass in order to reduce the number of
parameters. Most of the results are given for the representative
case where the charge of one type of particle is twice the charge
of the other one. Special attention will be given to the
particular case where the charges are almost equal. In this regime
at zero temperature, an asymmetrical distribution of charges over
the chains was found previously \cite{wandprb08}, while here we
find that  temperature leads to a mixing of the particles and
finally a homogeneous distribution of particles.

The paper is organized as follows. In Sec. II, we describe the
model and the procedure used to calculate the melting temperature.
In Sec. III the results for the melting of different ordered
structures corresponding to different regimes are discussed. Our
conclusions are given in Sec. IV.

\section{Model}
\label{sec:model}

We study a two-dimensional binary cluster consisting of an equal
number of particles with distinct charges $q_a$ and $q_b$, which
are allowed to move in the $x$-$y$ plane. The charged particles
interact through a repulsive Debye-H\"{u}ckel (or Yukawa)
potential $exp(-r/\lambda)/r$, are free to move in the $x$
direction, and are confined by an external one-dimensional
parabolic potential which limits the motion of the particles in
the $y$ direction. A parabolic confinement potential was realized
in Ref. [\onlinecite{liu03}] for a $quasi$-one-dimensional system
of dusty plasma and in Ref. [\onlinecite{bubeck01}] for the case
of colloidal particles confined into a droplet region. Such a
Yukawa inter-particle potential describes, e.g. charged colloidal
particles in a liquid environment, and dusty particles in a
plasma.  The screening length $\lambda$ can be varied by changing
the density of the counterions (plasma) in the colloidal (dusty
plasma) system. In the case of a dusty plasma the lateral
parabolic confinement potential was recently realized by
structuring the bottom electrode into a channel \cite{melzer06}.
It was demonstrated that the electric field induced potential is
parabolic whose strength depends on the width of the channel and
the electric field strength. For the charged colloidal system the
colloids are e.g. confined between two glass plates which can be
structured into a channel \cite{haghgooie06,koppl06,haghgooie04}.
In this case the confinement potential can be represented by a
hard wall potential. It was shown recently \cite{bubeck98} that by
curving the bottom of the channel it is possible to induce a
parabolic confinement potential for the colloids where the
confinement frequency can be varied by changing the curvature.

The potential energy of the system is given by

\begin{figure*}
\begin{center}
\includegraphics[scale=0.75]{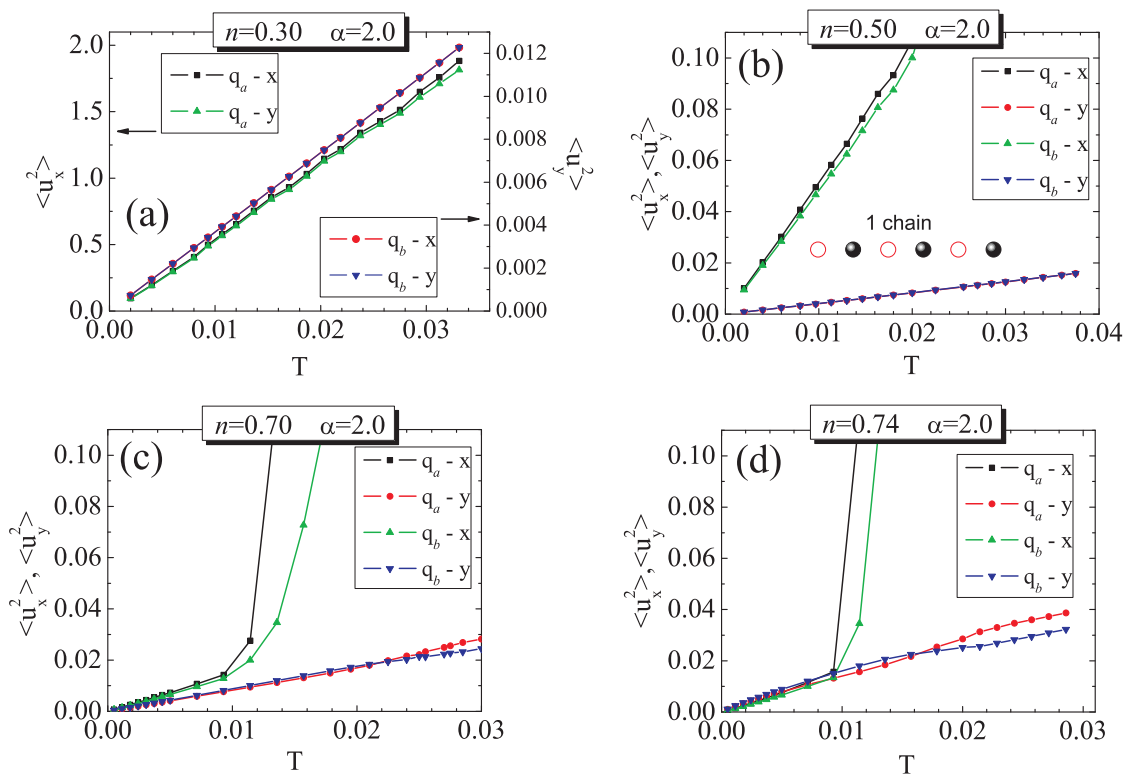}
\caption{(Color ) The relative displacements squared $\langle
u_x^{2} \rangle$ and $\langle u_y^{2} \rangle$ for the distinct
types of charges ($q_a=1$, $q_b=2$) as a function of temperature
for different densities (a) $n=0.30$, (b) $n=0.50$, (c) $n=0.70$,
and (d) $n=0.74$ in the single-chain regime. The ground state
configuration for the single-chain regime is shown as inset in
(b).}\label{fig:ur2_1chain}
\end{center}
\end{figure*}
\begin{equation} \label{eq:hamiltonian}
\begin{split}
  H = &\sum_{i}\frac{1}{2}m \omega_0^2 y_i^2 + \frac{q_{a}q_{b}}{\epsilon}\sum_{m}\sum_{n}
  \frac{exp(- \vert \mathbf{r}^{(a)}_{m} - \mathbf{r}^{(b)}_{n}\vert/\lambda)}{\vert \mathbf{r}^{(a)}_{m} -
  \mathbf{r}^{(b)}_{n}\vert}
   \\&  + \frac{q_{a}^{2}}{\epsilon} \sum_{i>j}\frac{exp(-\vert \mathbf{r}^{(a)}_{i} - \mathbf{r}^{(a)}_{j}\vert/\lambda)}{\vert \mathbf{r}^{(a)}_{i}
    - \mathbf{r}^{(a)}_{j}\vert}
\\& + \frac{q_b^2}{\epsilon} \sum_{k>l} \frac{exp(-\vert
\mathbf{r}^{(b)}_{k} - \mathbf{r}^{(b)}_{l}\vert/\lambda)}{\vert
\mathbf{r}^{(b)}_{k} - \mathbf{r}^{(b)}_{l}\vert},
\end{split}
\end{equation}
where $\epsilon$ is the dielectric constant of the medium the
particles are moving in, $\lambda$ is the Debye screening length,
and $\mathbf{r}^{(A)}_i=(x_i,y_i)$ is the distance of the $i^{th}$
particle of type $A$. In order to reveal the important independent
parameters of the system, it is convenient to write the energy and
the distances in units of $E_{0}=(m \omega
_{0}^{2}q_a^{4}/2\epsilon ^{2})^{1/3}$ and $r_{0}=(2q_a^{2}/m
\epsilon \omega _{0}^{2})^{1/3}$, respectively, and define the
quantity $\alpha = q_b/q_a$ (in the following we express $q_a$ and
$q_b$ in units of $e$, the elementary charge), and the screening
parameter $\kappa = r_{0}/\lambda$. In so doing, the expression
for the potential energy is reduced to
\begin{equation}\label{eq:hamiltonianII}
\begin{split}
  H = &\sum_{i}y_i^2+ \alpha \sum_{m}\sum_{n}
  \frac{exp(- \kappa\vert \mathbf{r}^{(a)}_{m} - \mathbf{r}^{(b)}_{n}\vert)}{\vert \mathbf{r}^{(a)}_{m} -
  \mathbf{r}^{(b)}_{n}\vert}
   \\ &+ \sum_{i>j}\frac{exp(-\kappa\vert \mathbf{r}^{(a)}_{i} - \mathbf{r}^{(a)}_{j}\vert)}{\vert \mathbf{r}^{(a)}_{i}
    - \mathbf{r}^{(a)}_{j}\vert}
   \\ &+ \alpha^{2} \sum_{k>l} \frac{exp(-\kappa\vert
\mathbf{r}^{(b)}_{k} - \mathbf{r}^{(b)}_{l}\vert)}{\vert
\mathbf{r}^{(b)}_{k} - \mathbf{r}^{(b)}_{l}\vert},
\end{split}
\end{equation}
and the state of the system is determined now by the ratio between
the charges $\alpha$, the number of particles per computational
unit cell (which is related to the linear density), and the
dimensionless screening length $\kappa$. The dimensionless linear
density $n$ is defined here as the ratio between the number of
particles in the unit cell and the length of the unit cell in the
unconfined $x$ direction. Periodic boundary conditions were
considered in the $x$ direction in order to mimic an infinite
system. The largest size for the simulation box that we considered
along the $x$ direction was $L=50$ unit cells (N=400 particles),
and the results were the same as the ones found with $L=25$ (N=200
particles) and $L=12.5$ (N=100 particles) unit cells, indicating
that the results do not depend on this length. The temperature is
expressed in units of $T_{0}=E_{0}/k_B$, where $k_B$ is the
Boltzmann constant. From now on, all quantities presented in the
manuscript are in dimensionless units. The charges, in particular,
is in units of the elementary charge $e$.

The minimum energy configurations and the normal modes of this system
were systematically analyzed recently in Ref.
[\onlinecite{wandprb08}]. In the present work we employ Monte Carlo
(MC) simulations to investigate the thermal dependence of the minimum
energy arrangements. Starting from a small temperature, typically
$5\times10^{-4}T_0$, the system is equilibrated during typically
$10^{6}$ MC steps. The properties of interest are then measured and
averaged over the next $10^{6}$ MC steps. This procedure was
typically repeated 15-20 times and the quantities of interest
averaged over those 15-20 runs. The temperature was increased in
steps of the order $\Delta T \approx 10^{-3} - 10^{-2}$, and the
previous process was repeated.

It is well-known that in 2D and 1D the root mean square deviation of
the particles from their average lattice position diverge with the
size of the system when the temperature is nonzero. This feature was
interpreted as preventing long-range order in 2D and 1D. In the
meantime we know that 2D and 1D ordering is possible even at nonzero
temperature. In order to characterize melting, we consider therefore
a modified Lindemann parameter \cite{bedanov94,gio04}, $L_p = \langle
u^{2} \rangle/d_r^2$, where $\langle u^{2} \rangle$ is the difference
between the mean square displacement of neighboring particles from
the respective equilibrium positions $\vec{r}_0$, and $d_r$ is the
mean interparticle distance which is defined below. The quantity
$\langle u^{2} \rangle$ is given by the expression
\begin{equation}\label{eq:ur2}
\begin{split}
\displaystyle{
  \langle u^{2} \rangle \equiv
  \frac{1}{N} \Big\langle \sum_{i=1}^{N}\frac{1}{N_{nb}}\sum_{j=1}^{N_{nb}}
  [ (\vec{r}_{i}-\vec{r}_{0i}) - (\vec{r}_{j}-\vec{r}_{0j})]^2 \Big\rangle,
}
\end{split}
\end{equation}
where $\langle\rangle$ means the average over the MC steps, $N$ is
the total number of particles in our simulation cell, and $N_{nb}$
is the number of neighbors of the i{\it th} particle. Previously,
it was shown that this modified Lindemann parameter characterized
the melting transition very well \cite{bedanov85}. The melting in
the $x$ and $y$ directions was characterized by the quantities
$\langle u_x^{2}\rangle/d_x^2$ and $\langle u_y^{2}
\rangle/d_y^2$, where the relevant distance in each direction,
$d_x$ and $d_y$, are respectively, given by the mean distance
between particles in each row, and the distance between chains. In
the single-chain regime we study the behavior of $\langle u_y^{2}
\rangle$ (not $L_{py}$), since there is no characteristic distance
in the $y$ direction. In the $x$ direction, the modified Lindemann
parameter $\langle u_x^{2} \rangle/d_x^2$ can be defined normally,
where $d_x$ is the average distance between particles along the
chain.

Besides the modified Lindemann parameter described above, we also
calculated other quantities in order to better characterize the
melting phenomenon. An example is the correlation function
$G_{\eta}$, given by:
\begin{equation}\label{eq:gtr}
\begin{split}
\displaystyle{
  G_{\eta}=
  \frac{1}{N} \Big{\langle} \sum_{i=1}^{N} \frac{1}{N_{nb}}\sum_{j=1}^{N_{nb}}
  \cos\Big(\frac{2 \pi (\eta_i - \eta_j)}{d_\eta}\Big) \Big\rangle,
}
\end{split}
\end{equation}
where $\eta=x,y$ and $d_\eta =d_x,d_y$ (defined above). The
function $G_{\eta}$ indicates how ordered a configuration is.
Specifically, it compares the distance between particles for a
given temperature with the distance between particles in the
ground state configuration ($T=0$). In the case of a highly
ordered configuration, $G_{\eta}$ is equal to unity. For the
completely disordered case the correlation function is zero. In
the present study, the $G_{\eta}$ function is calculated for each
type of particle separately. This means in the single- and
two-chain configurations it is defined only in the $x$ direction.

\section{Results}

In general, the present binary system can be characterized by an
asymmetrical thermal behavior with respect to the spatial
directions $x$ and $y$, and with respect to the two types of
particles. Anisotropy with respect to the spatial direction was
also found in the mono-disperse system \cite{gio04}, but we found
here that the presence of different types of charges introduces
new features. For some cases, a very restrictive thermal behavior
is observed only for one type of particle.

As a representative example we study the case in which the ratio
between charges is $\alpha=2$, and the screening parameter
$\kappa=1$, which is a typical value for colloids and dusty
plasmas \cite{bonitz06}. The thermal stability of the structures
will be analyzed for different densities. We will also pay special
attention to the case when the distinct types of charges are
almost equal ($\alpha \approx 1$). Earlier we found that in this
case, the particles are arranged in a mixed configuration where
both types of charges are randomly distributed over the rows
\cite{wandprb08}. The zero temperature phase diagram of the
present system can be found in Fig. 3 of Ref.
[\onlinecite{wandprb08}].

We start by presenting the results for the one-chain regime, where
the mean squared displacements have a very particular behavior as
a function of temperature. Next, the multi-chain regime will be
studied, where we concentrate on the two- and four-chain regimes.
For the parameters $\alpha=2$, $\kappa=1$, and the interval of
density considered here the system is found only in the minimum
energy structures with one, two, and four chains \cite{wandprb08}.
Finally, the interesting mixed or disordered case is investigated.

\subsection{The single-chain regime}
\label{sec:onechain}

\begin{figure}
\begin{center}
\includegraphics[scale=0.80]{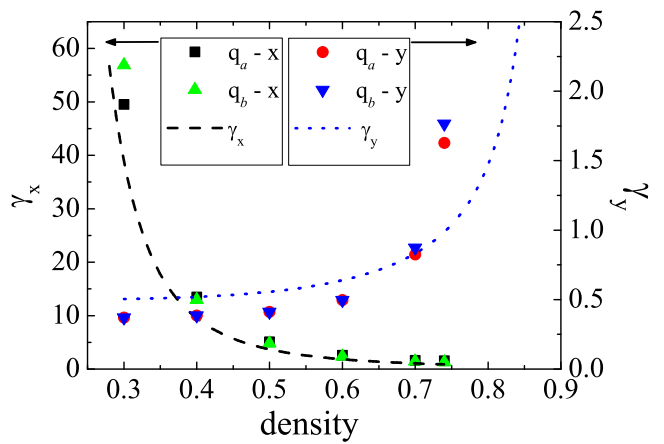}
\caption{(Color ) The slopes $\gamma_x$ and $\gamma_y$ in the low
temperature interval of the curves $\langle u_\eta^{2} \rangle$
{\it versus} $T$ ($\eta=x,y$) obtained from simulations in the
single-chain regime, and analytically as given by Eqs.
(\ref{eq:slopeX}) and (\ref{eq:slopeY}) as a function of
density.}\label{fig:slope_1chain}
\end{center}
\end{figure}

The single-chain regime is very peculiar. In principle, a
Lindemann-like criterion can not be applied here since for small
densities the difference in the mean square displacement behaves
linearly with temperature, and no clear melting transition is
found. For high densities, the usual abrupt increase of the mean
square displacement is found at some critical temperature which is
characteristic for melting, but this is only found along the
chain. In addition, the dimensionless modified Lindemann parameter
presented in Sec. \ref{sec:model} is  not defined in the $y$
direction, since there is no characteristic distance in this
direction. In this case, we still study the thermal behavior of
$\langle u_y^{2} \rangle$, without reference to a melting
temperature.

For $\alpha=2$ and $\kappa=1$ the one-chain configuration [see
Fig. \ref{fig:phase_diagram} in Appendix A] is found in the
interval of density $0<n\lesssim0.74155$. For small densities, $n
< 0.5$, the mean square displacements $\langle u_x^{2}\rangle$
(and also $L_{px}$) and $\langle u_y^{2} \rangle$ increase
linearly with temperature, typical for harmonic oscillations of
particles around their respective equilibrium positions [Fig.
\ref{fig:ur2_1chain}(a)]. The slopes of $\langle u_x^{2} \rangle$
and $\langle u_y^{2} \rangle$ are almost the same for the two
distinct charges, but they are very different along the $x$ and
$y$ directions, being a factor of almost 20 larger in the former.
The reason for such an asymmetry can be qualitatively understood
by the fact that a parabolic confinement in the $y$ direction is
equivalent to a homogeneous distribution of background charge in
that direction. This charge distribution is constant, independent
of temperature, and generates a constant pressure in the direction
perpendicular to the chain of charges. On the other hand, the
pressure along the chain direction is a consequence of the
interaction of each charge with all other particles in the chain.
Such an interaction is strongly affected by the relative positions
of the particles, i.e. the density, and increases with increasing
density. For small densities the restoring force to bring the
particles back to their equilibrium position is very small which
explains the large values obtained for $\langle u_x^{2} \rangle$.

\begin{figure}
\begin{center}
\includegraphics[scale=0.85]{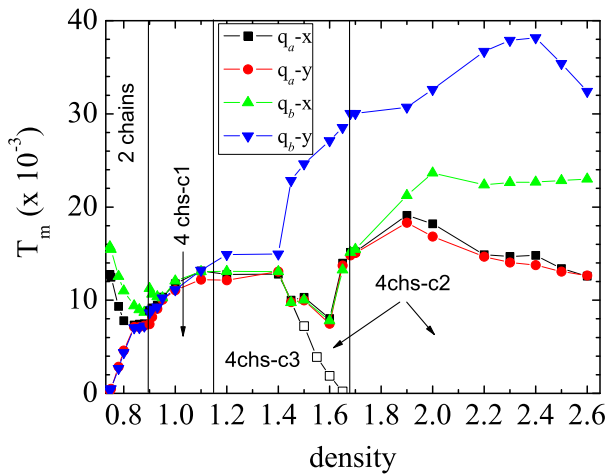}
\caption{(Color ) The melting temperature of the distinct types of
charges ($q_a=1$, $q_b=2$) and different directions ($x$, $y$) for
the multi-chain regime as a function of density ($n$). The
open-square symbols indicate the temperature for which a plateau
in the  $\langle u^{2} \rangle$ {\it versus} $T$ curves appear in
the four-chain (case 3) regime.}\label{fig:melt_temp}
\end{center}
\end{figure}

For $n\gtrsim0.5$ the behavior of the quantity $\langle u_y^{2}
\rangle$ remains linear with temperature, but the temperature
dependence of $\langle u_x^{2}\rangle$ is qualitatively different.
The harmonic oscillation (linear part) observed for small
temperatures is now followed by a nonlinear regime, which becomes
more abrupt for larger densities. As illustrated in Figs.
\ref{fig:ur2_1chain}(c-d), the temperature at which $\langle u_x^{2}
\rangle/d_x^2$ increases abruptly becomes smaller with increasing
density, and depends on the type of particle. For larger densities,
fluctuations of the particle positions can destabilize the ordered
structure more easily. This behavior is very different from the
melting of a classical 2D Wigner solid where the melting temperature
has the dependence $T_c \sim \sqrt{n}$.

Another qualitative modification of the $\langle u^{2} \rangle$ {\it
versus} $T$ curves with increasing density is related with the linear
regime. Notice that for large densities the slope of the linear part
of the $\langle u_x^{2} \rangle$ curves becomes smaller than the one
in the $y$ direction. This is a consequence of the fact that in the
$y$-direction there is an energy barrier for the zig-zag transition
\cite{gio04,zigcomment}, and the barrier height decreases with
increasing density and consequently the effective potential minima
become shallower.

An approximately analytic expression for the slope of the linear
$T$-dependence of $\langle u_\eta^{2} \rangle$ can be obtained .
Consider the potential energy given by Eq. (\ref{eq:hamiltonianII}),
next we limit ourselves to the nearest neighbors and then expand the
Yukawa interaction terms with respect to small oscillations. In this
way we obtain the deviation of the particles, within the harmonic
approximation, from which we obtain expressions for $\langle
u_\eta^{2} \rangle$. Within the harmonic approximation we have
$\langle u_{\eta}^{2} \rangle=\frac{1}{2}k_B T$ from which we found,
in dimensionless units, the slopes
\begin{equation}\label{eq:slopeX}
  \gamma_x \approx [4\alpha n^2 \kappa\exp(-\kappa/n)]^{-1} ,
\end{equation}
and
\begin{equation}\label{eq:slopeY}
  \gamma_y \approx [2-2\alpha n^2 \big(n+\kappa
  \big)\exp(-\kappa/n)]^{-1},
\end{equation}

\noindent where the density in the single-chain regime is $n=1/a$,
with $a$ the distance between neighbor charges along the chain.
From Eqs. (\ref{eq:slopeX}) and (\ref{eq:slopeY}) we note that the
slope $\gamma_y$ increases with density. In the $x$ direction, the
opposite behavior is found for $\gamma_x$. These approximate
analytic expressions are compared with the results from our
simulations in Fig. \ref{fig:slope_1chain}. In spite of the rough
approximation used to obtain Eqs. (\ref{eq:slopeX}) and
(\ref{eq:slopeY}), there is good qualitative agreement with the
numerical results, especially in the $x$ direction, which
corroborates with the qualitative explanation commented before.
Note that from Eq. (\ref{eq:slopeY}) we find that $\gamma_y$
diverges for $n \approx 0.897$ which signals the zig-zag
transition and compares with the value $n=0.74155$ which we found
previously from our MC calculation. The difference is due to
non-linear effects and contributions of particles beyond the
nearest neighbor.

\subsection{The multi-chain regime}
\label{sec:multichain}

\begin{figure}
\begin{center}
\includegraphics[scale=0.65]{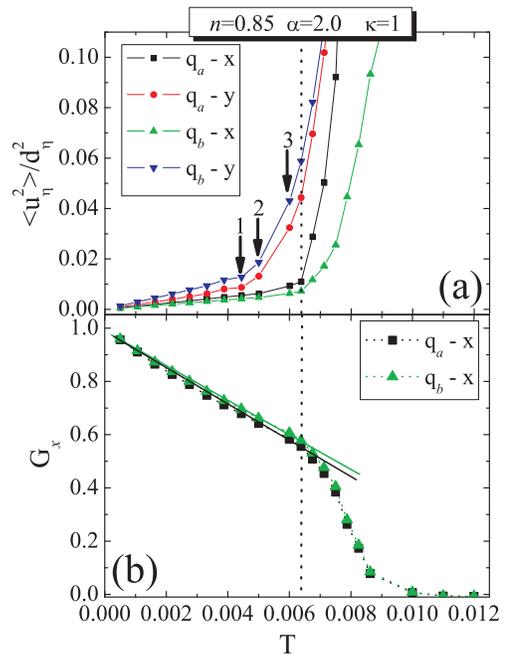}
\caption{(Color ) (a) The Lindemann parameter $\langle
u_{\eta}^{2} \rangle/d_\eta^2$ ($\eta=x,y$) for charges $q_a=1$
and $q_b=2$), and (b) the correlation function $G_{x}$ as a
function of temperature for a system with density $n=0.85$,
$\alpha=2$ and $\kappa=1$ (two-chain configuration). The vertical
dotted lines cross the full lines at $G_x =
0.5$.}\label{fig:n085_2chains_ur2}
\end{center}
\end{figure}

In this subsection the thermal behavior for the two- and
four-chain structures are presented (Note that a three-chain
structure is not stable; see Fig. \ref{fig:phase_diagram} in
Appendix A). According to Ref. [\onlinecite{wandprb08}], the
latter phase can be found in two different configurations, defined
as case 1 and case 2. However, we find here that a new four-chain
(case 3) configuration is possible between these two phases. The
new phase diagram is shown as Fig. \ref{fig:phase_diagram} of
Appendix A and replaces Fig. 3 of Ref [\onlinecite{wandprb08}].

A general feature of the multi-chain regimes is an asymmetrical
melting in both the direction and the distinct types of particles. In
the two- and four-chain (case 2) regimes particles with smaller
charge melt first. This is not the case for the four-chain (case 1)
system.

\subsubsection{The two-chain regime}
\label{subsec:twochains}

For $n = 0.74155$  the system undergoes a continuous structural phase
transition from the one- to the two-chain configuration
\cite{wandprb08}. For the two-chain configuration, the melting is
anisotropic with respect to the charges and with respect to the
spatial directions $x$ and $y$. The corresponding melting
temperatures are summarized in Fig. \ref{fig:melt_temp}. As can be
seen in Fig. \ref{fig:melt_temp} for both types of particles, the
melting temperature increases with increasing density in the $y$
direction, while the opposite trend is found in the unconfined $x$
direction. The explanation for such a different dependence is
associated with the fact that the repulsive Coulomb interaction
increases when the density increases. Such an increase in the Coulomb
repulsion leads to a dramatic different behavior in the $x$ and $y$
directions. In the latter, and in association with the external
confinement potential, the amplitude of the oscillations are smaller,
which explains the behavior of the melting temperature in the $y$
direction [Fig. \ref{fig:melt_temp}]. The equilibrium along the $x$
direction is established by the balance of the electrostatic force
between charges. Thermal fluctuations of the particle positions
destroy such balance and, consequently, the ordered arrangement. A
larger density brings the charges closer to one another, increasing
the repulsive interaction. In this case, small fluctuations of the
particle positions (small temperature) are enough to destabilize the
ordered structure, and to melt the system in the unconfined
direction. This different behavior along the $x$ and $y$ directions
is in line with the results found for a single chain as presented in
Fig. \ref{fig:slope_1chain}.

\begin{figure}
\begin{center}
\includegraphics[scale=0.6]{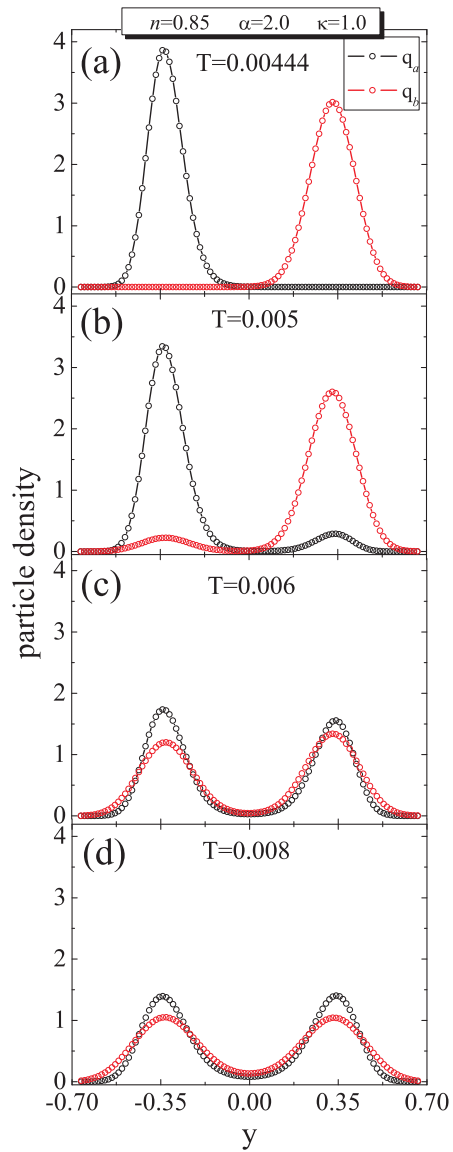}
\caption{(Color ) The distribution of particles along the confined
direction in the two-chain regime ($n=0.85$, $\alpha=2$ and
$\kappa=1$), for different temperatures (a) $T=0.00444$, (b)
$T=0.005$, (c) $T=0.006$, and (d)
$T=0.008$.}\label{fig:n085_2chains_ydens}
\end{center}
\end{figure}

In Fig. \ref{fig:melt_temp} we see that for a given density, we
find in the two-chain regime a smaller melting temperature in the
$y$ direction. Along the unconfined direction, the melting
temperature is different for both types of particles, being larger
for the bigger charge. This result is obtained according to the
modified Lindemann criterion ($\langle u_{\eta}^{2}
\rangle/d_\eta^2=0.1$), as illustrated in Fig.
\ref{fig:n085_2chains_ur2}(a) for a system with density $n=0.85$.
This conclusion is corroborated by our study through the
correlation function $G_{x}$ defined by Eq. (\ref{eq:gtr}). As
shown in Fig. \ref{fig:n085_2chains_ur2}(b), similar qualitative
and quantitative behavior is found for both charges along the
unconfined direction, where the $G_x$ value for the $a$-particles
is slightly smaller which is indicative for a lower melting
temperature.

In Fig. \ref{fig:n085_2chains_ydens}, we present the distribution
of particles along the $y$ direction for the same system (density
$n=0.85$), and for different temperatures. Three of those
temperatures, namely $T=0.00444$, $T=0.005$, and $T=0.006$, are
indicated in Fig. \ref{fig:n085_2chains_ur2}(a) by arrows $1$,
$2$, $3$, respectively. For all temperatures the distribution of
particles is asymmetric along the $y$ direction, and particles
with larger charge are more spread out  along the confined
direction, which is due to the larger Coulomb repulsion. For small
temperature (e.g., $T=0.00444$) the distinct types of particles
are segregated in different chains [Fig.
\ref{fig:n085_2chains_ydens}(a)], while with increasing
temperature the two types of particles become more and more mixed
over both chains [Fig. \ref{fig:n085_2chains_ydens}(d)].

The results shown in Fig. \ref{fig:n085_2chains_ydens} indicate
that the melting in the $y$ direction is not an abrupt process.
For $T=0.005$ [arrow 2 in Fig. \ref{fig:n085_2chains_ur2}(a)],
particles already start to jump between chains, which does not
occur in the linear regime of $\langle u_{y}^{2} \rangle/d_y^2$
[arrow 1 in Fig. \ref{fig:n085_2chains_ur2}(a)]. Note that for
$T=0.006$ [arrow 3 in Fig. \ref{fig:n085_2chains_ur2}(a)], which
is smaller than the melting temperature given by the
Lindemann-like criterion, particles of both chains are already
mixed, suggesting that the melting process in the $y$ direction is
not abrupt, and starts with a non-linear increase of the mean
square displacement. We notice from Fig.
\ref{fig:n085_2chains_ur2}(a) that $\langle u_{yi}^{2} \rangle$ is
a rather continuous function while $\langle u_{xi}^{2} \rangle$
exhibits a very steep increase at a certain temperature.

Along the chain direction, the temperature for which a deviation
from the linear decay of $G_{x}$ occurs [Fig.
\ref{fig:n085_2chains_ur2}(b)], coincides with the one along the
$x$ direction for which the Lindemann parameter deviates from a
linear temperature behavior [Fig. \ref{fig:n085_2chains_ur2}(a)].
Note that in this case, the deviation of the Lindemann parameter
from the linear regime starts at the same temperature ($T=0.007$)
for both types of charges [Fig. \ref{fig:n085_2chains_ur2}(a)] as
observed in Fig. \ref{fig:n085_2chains_ur2}(b) for $G_{x}$.  In
the unconfined direction, the correlation function together with
$\langle u_{xi}^{2} \rangle$ suggests a continuous disordering of
the system with increasing temperature.

\subsubsection{The four-chain (case 1 and case 3) regimes}
\label{subsec:fourchains_c1}

\begin{figure}
\begin{center}
\includegraphics[scale=0.4]{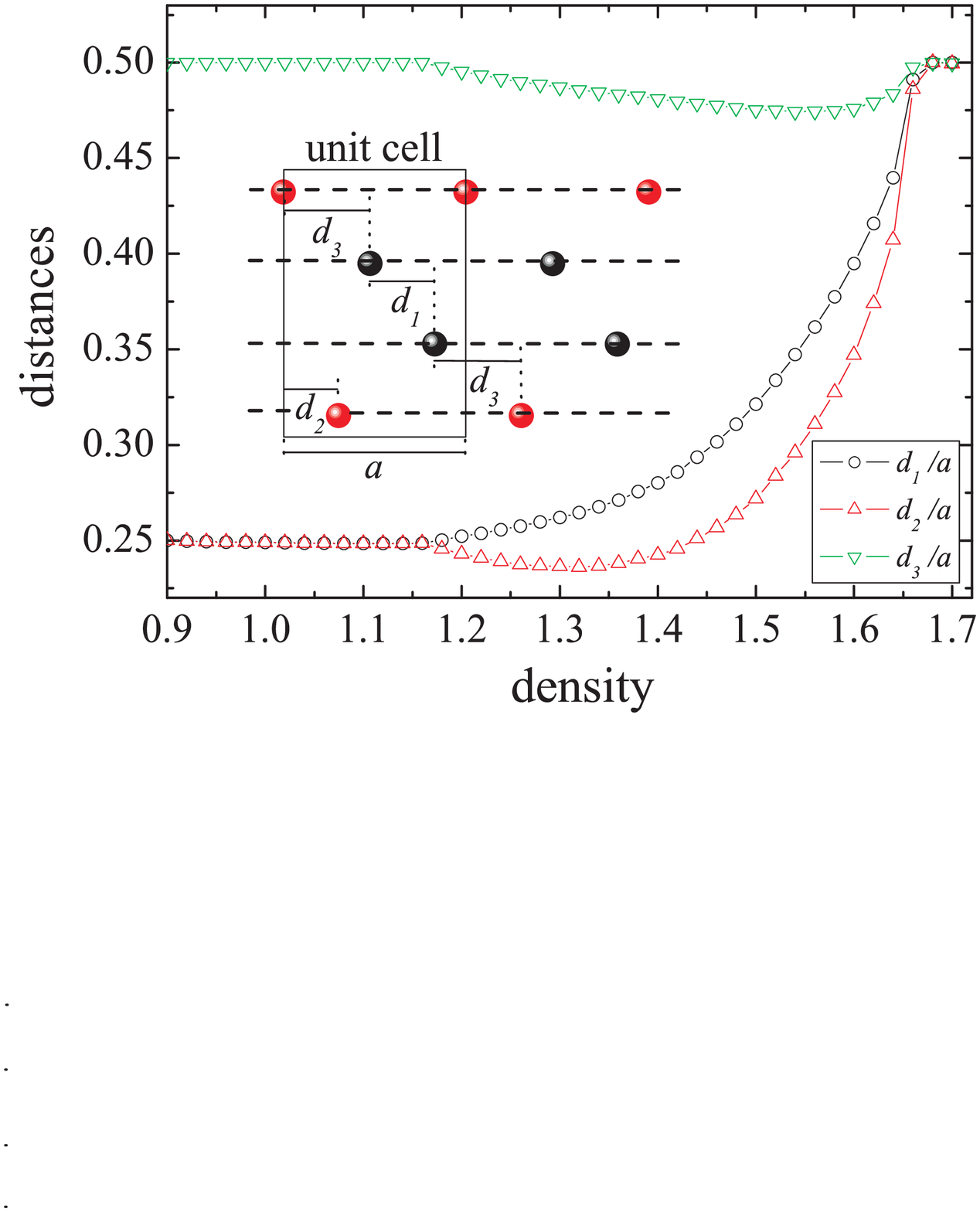}
\caption{(Color) The distances $d_1/a, d_2/a, d_3/a$, which
characterize the four-chain (case 3) regime, as a function of
density for a system with $\kappa=1$ and $\alpha=2$. A sketch of
the ground state configuration is presented as inset. $a$ is the
mean distance between particles along each chain.
}\label{fig:distances_case3}
\end{center}
\end{figure}
\begin{figure}
\begin{center}
\includegraphics[scale=0.75]{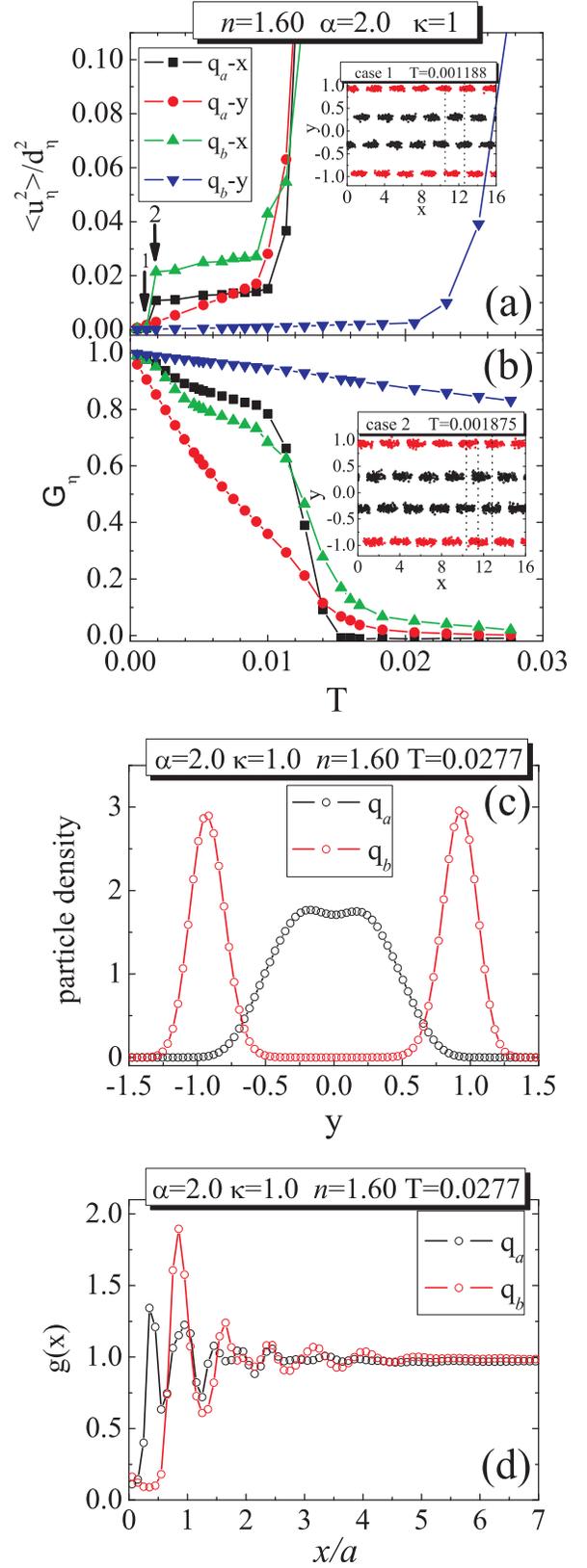}
\caption{(Color) (a) The Lindemann parameter $\langle u_{\eta}^{2}
\rangle/d_\eta^2$, and (b) the correlation function $G_{\eta}$
($\eta=x,y$) for charges $q_a=1$ and $q_b=2$ as a function of
temperature for a system with density $n=1.60$, $\alpha=2$ and
$\kappa=1$. Particle trajectories over 100 MC steps for
temperatures $T=0.001188$ (arrow $1$) and $T=0.001875$ (arrow $2$)
are shown as insets in (a) and (b), respectively. Black (red)
circles represent particles with charge $q_a=1$ ($q_b=2$). (c) The
distribution of particles in the $y$ direction. (d) The
pair-correlation function along the $x$ direction. $a$ is the mean
distance between particles along each chain in the ground state
configuration. }\label{fig:n160_4chc1_ur2}
\end{center}
\end{figure}

A further increase in the density brings the system to the
four-chain regime, which is found in three different
configurations. In all of them (case 1, case 2, case 3), the
different types of particles are symmetrically distributed with
respect to the $x$-axis, but segregated in distinct chains. In
this section we discuss the thermal behavior of case 1 and case 3.

In the four-chain (case 1) regime, which is found in the density
interval $0.9 \lesssim n \lesssim 1.16$, chains with equal charges
are displaced with respect to each other by $a/4$ ($a$ is the
distance between particles in each chain) along the chain
direction ($x$-axis), while neighbor rows with distinct types of
particles are displaced by a distance $a/2$ along the $x$-axis.
The distance between the internal chains (consisting of particles
with the same charge) is larger than the distance between the
internal chains and the external ones \cite{wandprb08}. This is
interesting because the interaction between chains with distinct
charges is intuitively expected to be larger than the interaction
between chains with the same lower charge. In the case 1 regime
the melting temperature in both spatial directions and for the two
different particles is almost the same as shown in Fig.
\ref{fig:melt_temp}.

For $1.16 \lesssim n \lesssim 1.68$ the system suffers a
continuous transition to the four-chain (case 3) regime which was
not predicted in Ref. [\onlinecite{wandprb08}]. Details of the
internal configuration of particles in such a regime is presented
as inset in Fig. \ref{fig:distances_case3}. The main difference
with respect to the case 1 regime is the density dependence of the
distances ($d_1,d_2,d_3$) which characterizes the ground state
configuration. In the four-chain (case 1) regime $d_1 =d_2 = a/4$,
and $d_3 =a/2$ ($a$ is the distance between particles along each
chain). As will be presented in the next section, the four-chain
(case 2) configuration is characterized by $d_1 =d_2 = d_3= a/2$.

 The behavior of the melting temperature in the four-chain (case
3) regime is rather distinct when compared to the one in case 1.
There is a sudden increase of the melting temperature of the
external chains in the $y$ direction for $n>1.4$. On the other
hand, the melting temperature for the internal chains decreases
with increasing density. The same decrease in the melting
temperature is observed for the external chains in the $x$
direction. The reason for the modified density dependence of the
melting temperature is a thermally structural phase transition in
which the system changes from the four-chain (case 3) to the
four-chain (case 2) configuration. For example, in Fig.
\ref{fig:n160_4chc1_ur2}(a) the Lindemann parameter $\langle
u_{\eta}^{2} \rangle/d_\eta^2$ ($\eta=x,y$) for charges $q_a$ and
$q_b$ is presented as function of temperature for a system with
$n=1.60$. As can be seen, there is a small plateau in the mean
square displacement for both charges, but only in the unconfined
direction. As observed previously in the case of circularly
confined finite size clusters \cite{wandpre05}, the plateau is a
typical signature of a thermally induced structural phase
transition, and we find here that the plateau is, in fact,
associated with such a transition. In the insets in Figs.
\ref{fig:n160_4chc1_ur2}(a) and \ref{fig:n160_4chc1_ur2}(b), we
present typical trajectories of the particles obtained over $100$
MC simulation steps for temperatures below and above the plateau
[indicated by the arrows $1$ and $2$ in Fig.
\ref{fig:n160_4chc1_ur2}(a)]. Around the structural phase
transition the system is in the harmonic regime, since the mean
square displacement increases linearly with temperature. For
temperatures before the plateau, the particles are arranged in the
four-chain (case 3) regime, while above the plateau the
trajectories of the particles indicate that the system is in the
four-chain (case 2) regime \cite{wandprb08}, whose features are
described in the next section. Such a thermally induced structural
phase transition is observed only for densities $n \gtrsim 1.50$
in the four-chain (case 3) regime. The temperature for which the
plateaus appears as a function of density is indicated in Fig.
\ref{fig:melt_temp} by the open square symbols. Above those
temperatures the system is found in the case 2 regime, as
indicated in Fig. \ref{fig:melt_temp}.

From the analysis of the Lindemann parameter, we find that the
thermal behavior depends on the spatial direction, which is
qualitatively similar for both types of particles. However, if the
correlation function $G_{\eta}$ ($\eta=x,y$) defined in Eq.
(\ref{eq:gtr}) is considered [Fig. \ref{fig:n160_4chc1_ur2}(b)],
we still observe a similar thermal dependence in the $x$ direction
for the charges $q_a$ and $q_b$. But this is different in the $y$
direction. The thermal dependence of $G_{y}$ for charge $q_a$
(internal chains) suggests a continuous loss of ordering with
increasing temperature, and differs from the one of particles with
charge $q_b$ (external chains). The latter thus indicates an
ordered structure due to the balance between the Coulomb repulsion
from the charges in the internal chains and the external pressure
from the parabolic confinement potential.

For high temperatures, e.g. $T=0.0277$, the internal chain-like
structure of the system is disappeared, while the external chains
still remain well defined [Fig \ref{fig:n160_4chc1_ur2}(c)].
However, for both internal and external chains a liquid-like
behavior is found along the chain direction. The mean distance
between particles with smaller charge ($q_a$) in the $x$ direction
becomes half the distance between particles with larger charge
($q_b$). This is also illustrated by Fig.
\ref{fig:n160_4chc1_ur2}(d), where the pair correlation function
$g(x)=\frac{L}{N^2}\sum_{i \neq
j}\langle\delta[x-(x_i-x_j)]\rangle$ is shown.

\begin{figure}
\begin{center}
\includegraphics[scale=0.65]{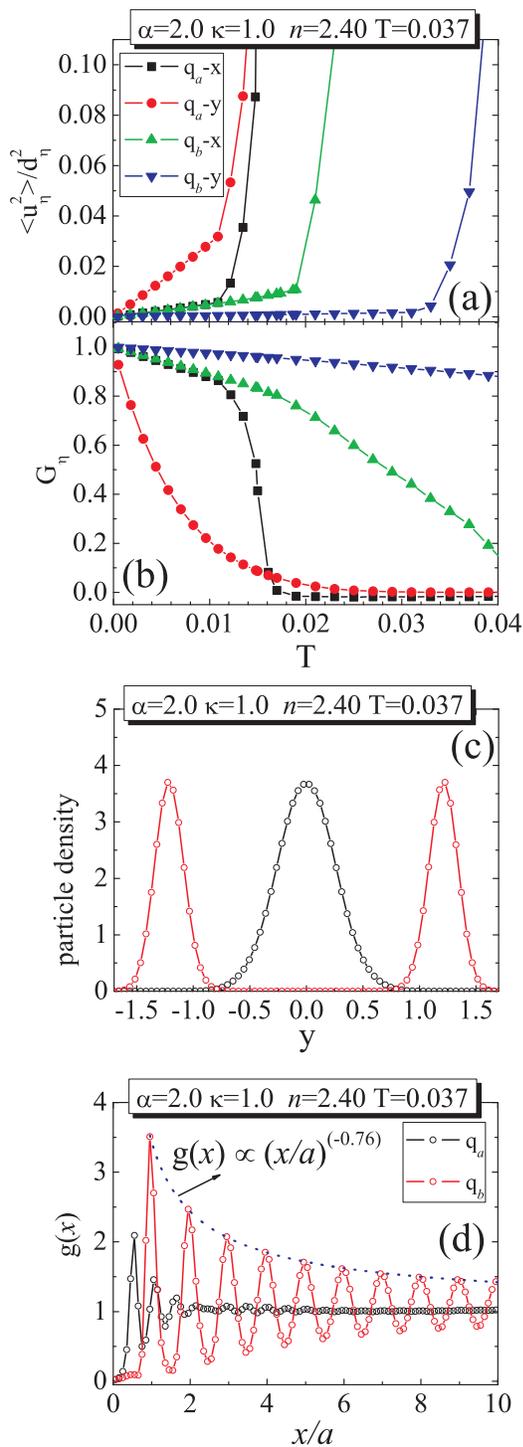}
\caption{(Color) (a) The Lindemann parameter $\langle u_{\eta}^{2}
\rangle/d_\eta^2$, and (b) the correlation function $G_{\eta}$
($\eta=x,y$) for charges $q_a$ and $q_b$ as a function of
temperature for a system with density $n=2.40$, $\alpha=2$ and
$\kappa=1$. (c) The distribution of particles in the $y$
direction. (d) The pair-correlation function along the $x$
direction. $a$ is the mean distance between particles along each
chain in the ground state configuration. The peaks of the pair
correlation function for the $b$-particles is fitted by an
algebraic decay represented by the dotted blue curve. The legend
in (b) is the same of (a), and the legend in (d) is the same of
(c).}\label{fig:n240_4chc2_all}
\end{center}
\end{figure}

\subsubsection{The four-chain (case 2) regime}
\label{subsec:fourchains_c2}

For $T=0$ there is a continuous structural phase transition, at
$n\approx 1.68$, from case 3 to case 2 (see Fig.
\ref{fig:distances_case3}). In the four-chain (case 2) regime,
chains with the same charge are displaced by $a/2$ with respect to
each other along the $x$ direction. The density dependence of the
distance between chains has an opposite behavior to the one of
case 1, i.e. the distance between chains with distinct charges is
larger than the distance between the internal rows
\cite{wandprb08}. The different internal structure of the
four-chain (case 2) regime leads to a different temperature
dependence for the mean square displacements. As can be observed
in Fig. \ref{fig:melt_temp}, the melting temperature for both
types of particles, and in the two directions has a different
behavior as a function of density in the two phases of the
four-chain regime.

The particular arrangement of the charges [see inset in Fig.
\ref{fig:distances_case3}] results in a larger distance between
the internal ($q_a$) and the external ($q_b$) chains than that
found in case 1. As a consequence, the external confinement
potential acts more strongly on the particles placed in the outer
rows, resulting in a larger melting temperature in the $y$
direction. Fig. \ref{fig:melt_temp} indicates a clear anisotropy
in the melting temperature of the external chains with respect to
the $x$ and $y$ directions.

Now, considering the inner chains we observe that particles in
different rows are already close such that fluctuations of the
particle positions are sufficient to destroy the ordering in both
directions. This fact occurs for a temperature smaller than those
needed to melt the external chains. We can conclude that according
to the modified Lindemann criterion considered in the present
paper, the melting in the four-chain (case 2) regime is
anisotropic with respect to both spatial directions and the
distinct types of charges.

We resort again to the correlation function $G_\eta$ ($\eta=x,y$)
and other quantities in order to better understand the temperature
dependence. As an example, in Figs. \ref{fig:n240_4chc2_all}(a)
and \ref{fig:n240_4chc2_all}(b), a comparison between the
temperature dependence of the Lindemann parameter and the
correlation function $G_{\eta}$ for density $n=2.40$ is made. In
both the $x$ and $y$ directions the Lindemann parameter has the
same qualitative behavior as a function of temperature for both
types of particles, i.e. a linear behavior followed by an abrupt
non-linear increase of $\langle u_{\eta}^{2} \rangle/d_\eta^2$
[Fig. \ref{fig:n240_4chc2_all}(a)]. On the other hand, $G_{\eta}$
presents a very different temperature dependence in the $x$ and
$y$ direction, and for both types of charges. As in the case 1
regime, $G_{y}$ shows a continuous loss of ordering as a function
of $T$ for the internal chains ($q_a$), which is not observed for
the external ones. In the $x$ direction, the behavior of the
internal and external chains is similar only for small
temperatures ($T \lesssim 0.01$). For $T \gtrsim 0.01$, the
behavior of $G_{x}$ for the internal chains exhibits an abrupt
decay in the corresponding temperature interval associated to the
melting according  to the Lindemann parameter criterion. For
particles in the external chains, we find a $G_y$ $T$-dependence
that suggests a slowly disordering with increasing temperature,
which is not seen from the Lindemann parameter [Fig.
\ref{fig:n240_4chc2_all}(a)]. For example, at the temperature
$T=0.037$ the Lindemann parameter indicates that the system is
completely melted, but $G_y$ indicates that only the internal
chains are completely disordered (i.e., $G_x \approx G_y \approx
0$).

In Figs. \ref{fig:n240_4chc2_all}(c) and
\ref{fig:n240_4chc2_all}(d), we show the distribution of particles
along the $y$ direction and the pair correlation function along
the unconfined direction for $T=0.037$, respectively. As can be
observed, the initial two-chain structure of the internal chains
is completely destroyed in the $y$ direction, while it is
relatively preserved in the external chains. Along the $x$
direction, Fig. \ref{fig:n240_4chc2_all}(d) indicates that the
internal chains are completely disordered, but the
pair-correlation function for particles in the external chains
presents an algebraic decay, which is indicative of long-range
order. Notice that the $G_x$ function for the $b$ particles still
presents a finite value for $T=0.037$ in agreement with the $g(x)$
result. For very high temperatures, and as found in the case 3
regime, the mean distance between the particles ($x$ direction)
with smaller charge becomes half the one for larger charge (first
peak in the pair correlation function), which indicates that a
disordered one chain is found in the internal part of the system.
The results presented here indicate a complete anisotropic
(direction and type of charge) melting for the binary system.

\begin{figure}
\begin{center}
\includegraphics[scale=0.75]{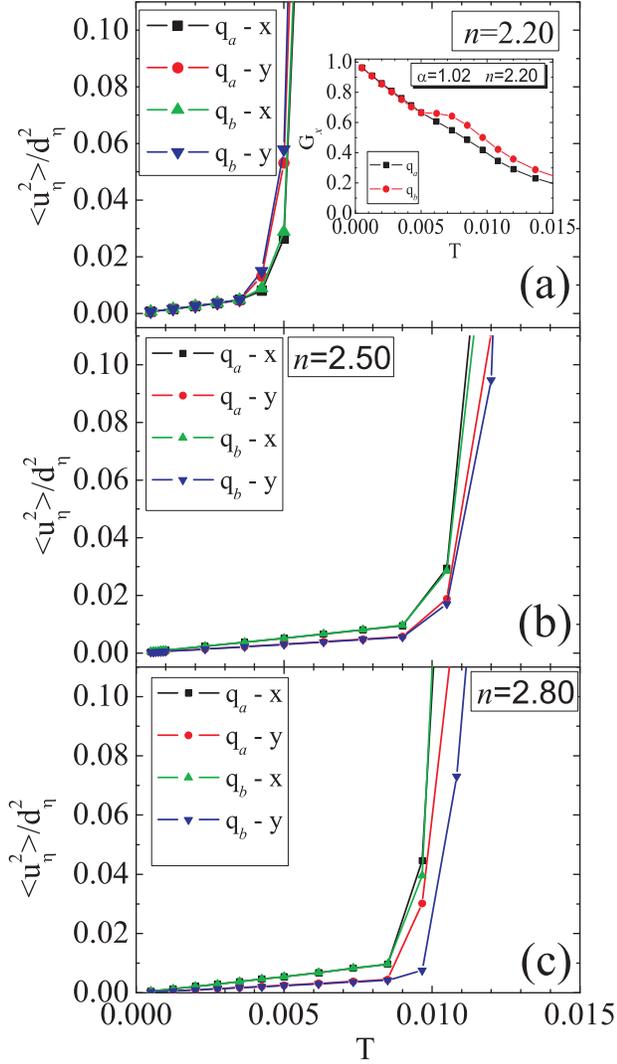}
\caption{(Color) The Lindemann parameter $\langle u_{\eta}^{2}
\rangle/d_\eta^2$ for a system in the mixed regime ($q_a=1$ and
$q_b=1.02$) as a function of temperature for different densities
(a) $n=2.20$, (b) $n=2.50$, and (c)
$n=2.80$.}\label{fig:alf1p02_ur2_mixing}
\end{center}
\end{figure}

\subsection{Mixed regime}

\begin{figure}
\begin{center}
\includegraphics[scale=0.6]{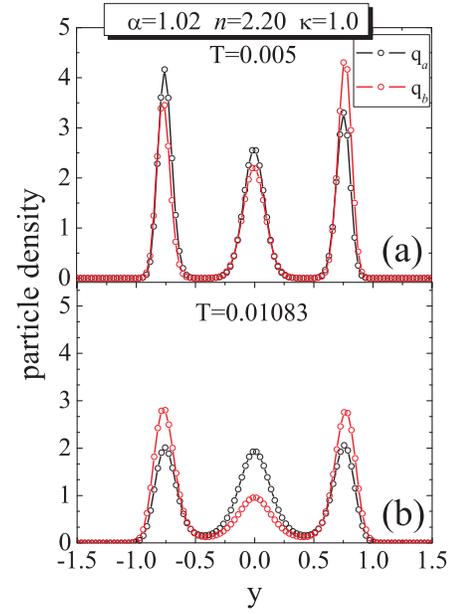}
\caption{(Color) The distribution of particles ($q_a=1$ and
$q_b=1.02$) along the confined $y$ direction for a system in the
mixed regime for temperatures (a) $T=0.005$, and (b)
$T=0.01083$.}\label{fig:density_mixing}
\end{center}
\end{figure}

\begin{figure}
\begin{center}
\includegraphics[scale=0.7]{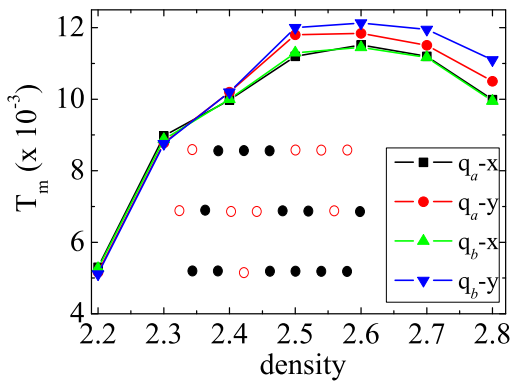}
\caption{(Color) The melting temperature of the distinct types of
particles ($q_a=1$ and $q_b=1.02$) as a function of density for a
system in the mixed regime. An example of the ground state
configuration is shown as inset}\label{fig:meltTemp_mixing}
\end{center}
\end{figure}

In the almost mono-disperse case (i.e. $\alpha \approx 1$), the
minimum energy arrangement is the three-chain configuration with
the charges randomly distributed over the chains. Neighbor chains
are displaced by a distance $a/2$ with respect to each other along
the unconfined direction [inset of Fig. \ref{fig:meltTemp_mixing}
]. For $\kappa=1$, the interval of density in which the mixed
regime can be observed depends on the ratio between the size of
the charges $\alpha$ \cite{wandprb08}. Here, we focus on the
representative case $\alpha=1.02$, i.e. particles have a
difference in their charges of $2 \%$ where, for $T=0$, the
disordered state was found in the density interval $2.14 \lesssim
n \lesssim 3.23$.

The temperature dependence of the Lindemann parameter is presented
in Fig. \ref{fig:alf1p02_ur2_mixing} for systems with
$\alpha=1.02$, $\kappa=1$, and densities $n=2.20, 2.50, 2.80$. In
spite of the small density interval in which the mixed regime is
found \cite{wandprb08}, a different qualitative behavior is
observed with increasing density. In general, the melting
temperature increases with increasing density. In addition, the
melting temperature in the $y$ direction becomes distinct for both
types of particles. In the $x$ direction, the melting temperature
depends on the density, but it is the same for the two set of
particles.

The random arrange in the mixed regime generates an unbalanced
distribution of the different charges over the chains, as can be
seen in Fig. \ref{fig:density_mixing}(a), where the distribution
of particles along the confined direction is shown for the system
with density $n=2.20$ ($\alpha=1.02$, $\kappa=1$) at $T=0.005$. An
interesting feature is observed when the temperature is increased.
In this case, a more homogeneous distribution of both types of
particles is induced, where particles with larger charge are
located predominantly at the edge chains [Fig.
\ref{fig:density_mixing}(b)]. A similar behavior is found for all
densities in the mixed regime. It is interesting to comment on the
more homogeneous distribution of particles which can also be
noticed from the correlation function. In particular, we present
as inset in Fig. \ref{fig:alf1p02_ur2_mixing}(a) the correlation
function $G_x$ as a function of temperature for the system with
$n=2.20$ ($\alpha=1.02$, $\kappa=1$). For $T \gtrsim 0.005$ the
correlation function $G_x$ is larger for particles with larger
charge ($q_b =1.02$), which coincides with the rapid increase of
$\langle u_y^{2} \rangle$ at which particles start to jump between
chains leading to a more homogeneous distribution of the different
particles over the chains. At this point the system melts. The
density dependence of the different melting temperatures of the
system with $\alpha=1.02$ and $\kappa=1$ in the mixed regime is
summarized in Fig. \ref{fig:meltTemp_mixing}.

\section{Conclusion}

We studied melting and thermal induced transitions of a
two-dimensional binary system consisting of an equal number of
particles with different charges interacting through a screened
Coulomb potential, and confined along one direction by a parabolic
potential. The different phases (see Fig. \ref{fig:phase_diagram})
that nucleate as a function of density \cite{wandprb08} exhibit a
very distinct and nontrivial temperature dependence.

The melting temperature was determined from a modified Lindemann
criterion \cite{gio04}. Other quantities as the pair-correlation
function, the distribution of particles, and the translational
correlation function were in addition studied to corroborate the
obtained values for the melting temperature and to enhance our
understanding of the melting process of the different ordered
structures.

We found that melting is anisotropic with respect to the spatial $x$
and $y$ directions, and with respect to the distinct type of
particles. In the multi-chain regime, depending on the internal
structure of the system, a very different temperature behavior is
found. Specifically, in the new four-chain (case 3) regime we found a
thermally induced structural phase transition, which shows up by the
appearance of a plateau in the curves of the mean square displacement
$versus$ temperature. For the four-chain (case 2) regime an analysis
of the pair-correlation function shows that the external chains,
composed by particles with larger charge, exhibits a relatively
ordered structure along the unconfined direction, even for a
temperature larger than the melting temperature as determined by the
Lindemann-like criterion. The correlation function, indicates a
qualitatively distinct melting in the $x$ and $y$ directions and for
the distinct types of particles.

To summarize, the two- to four-chain transition was found previously
to be a first order transition (at $T=0$)) and we found here that the
density at which the transition occurs is independent of temperature.
The different melting temperatures exhibit small jumps at this phase
transition. This is different for the different four-chain regimes,
which was previously found to crystallize only in two configurations
with a first order transition between them. Here we predict that a
new four-chain phase exist intermediate between the previous two
four-chain configurations that we studied in Ref.
[\onlinecite{wandprb08}]. As a consequence, the transitions between
the different four-chain phases are now continuous, i.e. second
order, instead of first order transitions. In addition, the density
at which the transition takes place depends on temperature which is a
qualitative difference from the mono-disperse case \cite{gio04}.

For $\alpha \approx 1$, the system can be found in a mixed
configuration which we studied in the regime of three chains. The
configuration is characterized by a random distribution of the two
types of charges over the chains. In this regime, we found that an
increase of temperature results in a more symmetric distribution
of particles over the different chains, and an increase of the
melting temperature with density, except for high densities.

The predicted melting phase diagram for the binary
$quasi$-one-dimensional system can be investigated experimentally
in systems as e.g. dusty plasmas and colloids. Previously,
harmonic confinement potentials have been realized in dusty plasma
\cite{liu03} and in colloidal \cite{bubeck98} systems. The phase
diagram can be scanned experimentally by changing e.g. the
density. Furthermore, in a dusty plasma it is also possible to
vary the effective temperature which is not possible for a
colloidal system where the particles move in a liquid environment.
But for the latter system one can also use paramagnetic particles
which, in the presence of a perpendicular magnetic field, become
magnetized resulting in a dipole inter-particle interaction
potential. The strength of this potential can be tuned by the
external magnetic field \cite{koppl06}. Because the obtained
results are generic, and qualitatively do not depend on the
specific functional form of the repulsive interaction potential,
similar melting behaviors are predicted for such a system of
dipoles. Because in classical melting it is the thermal energy
versus the average interaction energy that is the relevant
parameter the temperature scale in Fig. 3 can also be replaced by
the inverse of the interaction energy.

\section{acknowledgments}
This work was supported by CNPq, the Flemish Science Foundation
(FWO-Vl) and the bilateral program between Flanders and Brazil.

\appendix

\section{}
\label{app:appendix}

\begin{figure}
\begin{center}
\includegraphics[scale=0.8]{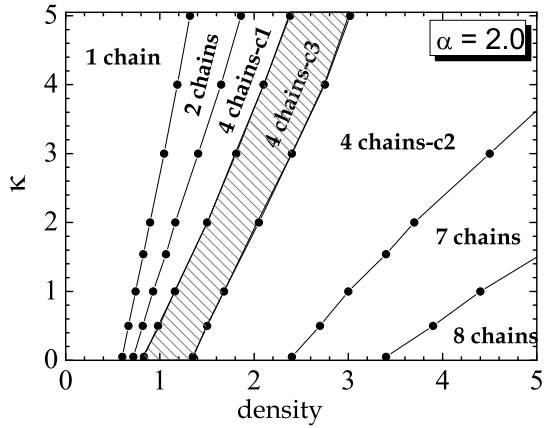}
\caption{(Color) The zero-temperature $\kappa$-density phase
diagram for $\alpha=2$. The new stability region for the
four-chain configuration is hatched.}\label{fig:phase_diagram}
\end{center}
\end{figure}

The zero-temperature phase diagram for $\alpha=2$ is shown in Fig.
\ref{fig:phase_diagram}, where we included the new four-chain
(case 3) regime. A sketch of the four-chain (case 3) configuration
is shown as inset in Fig. \ref{fig:distances_case3} and the
parameters characterizing this configuration are plotted in Fig.
\ref{fig:distances_case3} for $\kappa=1$ and $\alpha=2$. The phase
transitions between the different four-chain configurations is of
second order.

\end{document}